\documentclass[reprint]{revtex4-1}
\usepackage{amsmath}
\usepackage{amssymb}
\usepackage{graphicx}
\usepackage{bm}

\begin{document}
\title{The Effects of Turbulence on Three-Dimensional Magnetic
  Reconnection at the Magnetopause}

\author{L.~Price}
\affiliation{IREAP, University of Maryland, College Park MD
  20742-3511, USA}
\author{M.~Swisdak}
\affiliation{IREAP, University of Maryland, College Park MD
  20742-3511, USA}
\author{J.~F.~Drake}
\affiliation{IREAP, University of Maryland, College Park MD
  20742-3511, USA}
\author{P.~A.~Cassak}
\affiliation{Department of Physics, West Virginia University,
  Morgantown, West Virginia 26506, USA}
\author{J.~Dahlin}
\affiliation{IREAP, University of Maryland, College Park MD
  20742-3511, USA}
\author{R.~E.~Ergun}
\affiliation{Department of Astrophysical and Planetary Sciences,
  University of Colorado, Boulder, Colorado, 80303, USA}

\begin{abstract}

Two- and three-dimensional particle-in-cell simulations of a recent
encounter of the Magnetospheric Multiscale Mission (MMS) with an
electron diffusion region at the magnetopause are presented.  While
the two-dimensional simulation is laminar, turbulence develops at
both the x-line and along the magnetic separatrices in the
three-dimensional simulation. The turbulence is strong enough to make
the magnetic field around the reconnection island chaotic and produces
both anomalous resistivity and anomalous viscosity. Each contribute
significantly to breaking the frozen-in condition in the electron
diffusion region.  A surprise is that the crescent-shaped features in
velocity space seen both in MMS observations and in two-dimensional
simulations survive, even in the turbulent environment of the
three-dimensional system. This suggests that MMS's measurements of
crescent distributions do not exclude the possibility that turbulence
plays an important role in magnetopause reconnection.

\end{abstract}
\maketitle
%\begin{article}

\section{Introduction}

Magnetic reconnection facilitates the conversion of magnetic energy to
high-speed plasma flows and thermal energy. This energy release
requires a change in the topology of the field, which occurs at
magnetic x-lines.  Electron diffusion regions (EDRs), which surround
x-lines, are small, with characteristic thicknesses given by the
electron skin depth $d_e = c/\omega_{pe}$, where $\omega_{pe}$ is the
electron plasma frequency.  The detection of EDRs is the prime
motivation for the Magnetospheric Multiscale Mission (MMS).

The first stage of MMS has focused on the magnetopause, where the
differences between magnetospheric and magnetosheath plasma produce
asymmetric reconnection.  Some distinctive features of electron
distribution functions associated with the EDRs of asymmetric
reconnection have been recently identified
\citep{hesse14a,burch16a,bessho16a,shay16a}.  In particular, the
strong asymmetry in density across the magnetopause causes a large
component of the electric field perpendicular to the current sheet,
$E_N$, to form which, in turn, prevents the high-density magnetosheath
ions from crossing the magnetopause. (In the $LMN$ coordinate system
$L$ is in the direction of the reconnecting magnetic field, $N$
parallels the inflow direction and $M$ is perpendicular to $L$ and $N$
in the out-of-plane direction.)  This $E_N$ accelerates the
unmagnetized electrons near the magnetic null toward the
magnetosphere, where they are turned by $B_L$ into the $M$
direction. The result is cusp-like electron orbits on the earthward
side of the x-line and along the separatrices on the earthward edges
of the reconnection exhaust.  The consequence is crescent-shaped
velocity-space distributions which were first noted in numerical
simulations \citep{hesse14a}.  Parallel electric fields downstream
from the x-line also produce crescents along the outflow direction and
were first identified in the MMS data \citep{burch16a}.

Most simulations of reconnection treat a reduced geometry in which
variations in the out-of-plane direction are ignored.  This treatment
eliminates fluctuations with wavevectors in the invariant direction
and hence greatly inhibits the development of turbulence, which is
typically driven by the strong out-of-plane current in the diffusion
region.  Reconnection in this limit is essentially laminar, although
current-driven instabilities along the separatrices can produce
intense parallel electric fields \citep{cattell05a,lapenta11a}.  Since
crescents are only observed in regions of large $E_N$ where the
electron out-of-plane current $J_M$ is also large, the turbulence that
is expected to develop within these current layers might plausibly
scatter the electron orbits and destroy the crescents.  Thus, the fact
that crescent-shaped features are observed in MMS distribution
functions suggests that actual magnetopause reconnection is
laminar. The sensitivity of the electron crescents to the development
of turbulence remains an open issue.

A primary goal of MMS is to determine what breaks the frozen-in
condition during reconnection or, equivalently, what terms in Ohm's
law balance the out-of-plane reconnection electric field $E_M$ in the
EDR.  During asymmetric reconnection, the stagnation point of the
normal electron flow $v_{eN}$ is displaced toward the magnetosphere
side of the x-point \citep{cassak07a}. In two-dimensional simulations
the $M$ component of the divergence of the pressure tensor balances
$E_M$ at the stagnation point, but $E_M$ at the x-point can be
balanced by various terms (depending on the configuration), including
the electron inertia $m_ev_{eN}\partial v_{eM}/\partial N$
\citep{hesse14a}. An important question is whether the turbulence that
develops in the diffusion region alters these conclusions.

In this paper we present three-dimensional simulations of reconnection
with initial conditions reflective of the MMS event described in
\cite{burch16a}.  Because of the extra freedom associated with
dynamics in the dawn-dusk ($M$) direction, instabilities such as the
lower-hybrid drift instability (LHDI)
\citep{roytershteyn12a,pritchett11a,pritchett12a,pritchett13a} or the
electron Kelvin-Helmholtz instability \citep{lee15a} can develop.  In
contrast with the results of earlier simulations
\citep{roytershteyn12a,pritchett13a}, we find that for the parameters
associated with the MMS event, which has a larger jump in plasma
density than had been previously treated, the turbulence significantly
deforms the current layers and produces variations in the
electromagnetic fields sufficiently strong to affect the structure of
the diffusion region: anomalous resistivity and anomalous viscosity
both play a role in breaking the frozen-in condition.  (Interestingly,
high-frequency electric field fluctuations, amplitude $\gtrsim 20$
mV/m, were seen in the EDR during the MMS crossing.)  However, in
spite of the presence of turbulence in the simulations, crescents are
still present in the electron distribution functions within the strong
current layers on the magnetospheric edge of the diffusion region and
separatrices.  Thus, the role of turbulence in balancing Ohm's law
remains an open issue in the MMS observations.

\section{Simulations}\label{sims}

We use the particle-in-cell code {\tt p3d} \citep{zeiler02a}.  The
magnetic field strength $B_0$ and density $n_0$ define the Alfv\'en
speed $v_{A0}=\sqrt{B_0^2/4\pi m_in_0}$, with lengths normalized to the
ion inertial length $d_i =c/\omega_{pi}$, where $\omega_{pi}$ is the
ion plasma frequency, and times to the ion cyclotron time
$\Omega_{i0}^{-1}$.  Electric fields and temperatures are normalized
to $v_{A0}B_0/c$ and $m_iv_{A0}^2$, respectively.

The initial conditions for the simulations closely mimic those
observed by MMS during the diffusion region encounter discussed in
\cite{burch16a}.  The particle density $n$, reconnecting field
component $B_L$, and ion temperature $T_i$ vary as a function of $N$
with hyperbolic tangent profiles of width 1. The asymptotic values of
$n$, $B_L$, and $T_i$ are 1.0, 1.0 and 1.37 in the magnetosheath and
0.06, 1.70, and 7.73 in the magnetosphere.  The guide field
$B_M=0.099$ is initially uniform.  Pressure balance determines the
electron temperature $T_e$, subject to the constraint that its
asymptotic magnetosheath value is 0.12.  (The asymptotic value of
$T_e$ in the magnetosphere is thus 1.28.)  Although the system is in
force balance, the initial conditions are not an exact Vlasov
equilibrium.  Following initialization the system adjusts and reaches
the steady-state configuration analyzed here.

We performed both two-dimensional and three-dimensional simulations
with these parameters.  For the two-dimensional simulation the domain
had dimensions $(L_L,L_N) = (40.96,20.48)$ and employed the same
plasma parameters as that discussed in \cite{burch16a}.  The
three-dimensional simulation extended the $M$ direction:
$(L_L,L_M,L_N) = (40.96,10.24,20.48)$.  The ion-to-electron mass ratio
was set to $100$, which is sufficient to separate the electron and ion
scales.  The spatial grid has a resolution $\Delta = 0.02$ while the
smallest physical scale is the Debye length in the magnetosheath,
$\approx 0.03$.  As in \cite{burch16a} we used $500$ particles per
cell per species when $n=1.0$ for the two-dimensional simulation.  Due
to computational constraints, the three-dimensional simulation uses
$50$ particles per cell, which implies $\approx 3$ particles per cell
in the low-density magnetosphere.  To mitigate the resulting noise,
our analysis of this case employs averages over multiple cells.

The velocity of light is $c=15$ so that $\omega_{pe}/\Omega_{ce}=1.5$
in the asymptotic magnetosheath and 0.3 in the asymptotic
magnetosphere; the observed ratios are larger ($\approx 35$ and 6,
respectively).  As a result, the Debye length in the simulation is not
as small as at the magnetopause and might artificially suppress very
short wavelength electrostatic instabilities
\citep{jaraalmonte14a}. Unlike some earlier simulations of
\cite{roytershteyn12a} we do not force the rate of reconnection with
an external boundary condition; instead, the boundary conditions are
periodic in all directions. Our initial profiles also differ from
those earlier simulations (the density jump across the magnetopause
being 16 rather than 10) since they have been chosen to match the
event explored by MMS.

\begin{figure}
\includegraphics[width=0.9\columnwidth]{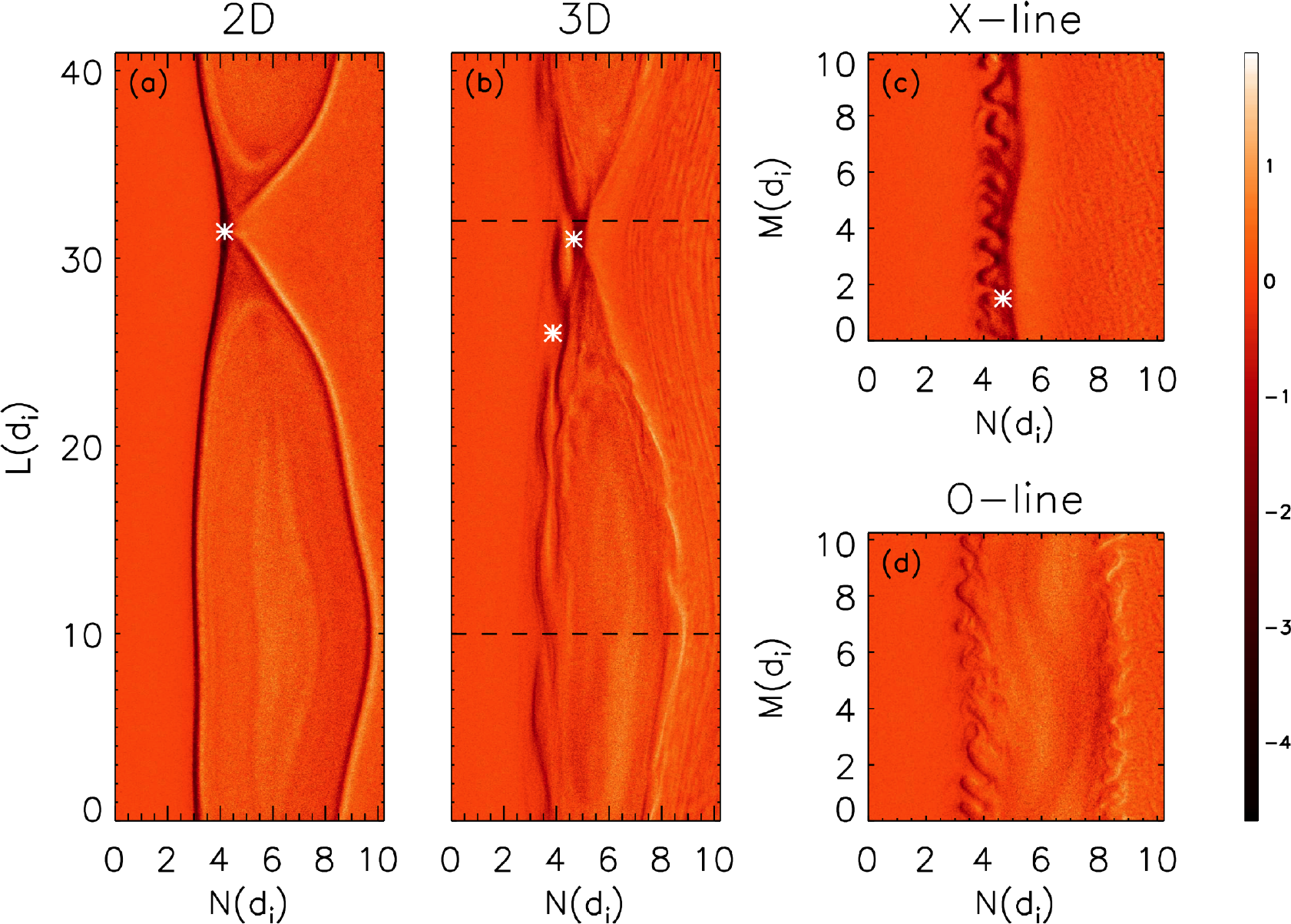}
\caption{\label{jez} Snapshots of $J_{eM}$, the dawn-dusk electron
  current density.  Panel (a): The $L-N$ plane from the
  two-dimensional simulation at $t=40$.  Panel (b): The $L-N$ plane
  from the three-dimensional simulation when roughly the same amount
  of flux has reconnected ($t=30$).  The two dashed lines denote the
  cuts shown in subsequent panels.  Panel (c): The $M-N$ plane from a
  cut through the x-line, the upper line in panel (b).  Panel (d): The
  $M-N$ plane from a cut through the island, the lower line in panel
  (b).  In each panel the colors are separately normalized; the bar at
  the right shows the relative variation.  The stars indicate the
  locations of the distribution functions presented in Figure
  \ref{cresdf}.}
\end{figure}

Figure \ref{jez} displays images of $J_{eM}$, the dawn-dusk electron
current density.  Panels (a) and (b) show the $L-N$ plane for the
two-dimensional and three-dimensional simulations after reconnection
of roughly the same amount of magnetic flux.  In both, the
magnetosphere (strong field, low density) is to the left and the
magnetosheath (weak field, high density) is to the right.  As is
typical in asymmetric configurations, the reconnection of equal
amounts of flux from the two sides means the islands bulge into the
magnetosheath.  While the two-dimensional simulation is laminar,
turbulence develops in the three-dimensional case.  This can be
clearly seen in panels (c) and (d), which show $J_{eM}$ in cuts
through the $M-N$ plane of the simulation at the locations denoted by
the dashed lines in panel (b).  The current layers at both the x-line
(panel c) and bordering the magnetic island (panel d) have become
turbulent.

The free energy in the strong, spatially localized, out-of-plane
electron flows are the likely drive for the instability.  The
wavelength is consistent with the lower-hybrid drift instability
(LHDI) both near the x-line and on the separatrices during asymmetric
reconnection.  The energy source for the LHDI is the relative drift of
the ions and electrons in the $M$ direction and the wavevector
satisfies the relation $\mathbf{k}\boldsymbol{\cdot}\mathbf{B}=0$ so
that $\mathbf{k}$ is along $M$ at the x-line and the midplane of the
island.  Thus, the LHDI does not develop in the two-dimensional
simulation.  Within the current layer, the range of excited
wavenumbers is relatively broad, $(m_e/m_i)^{0.25} \lesssim k\rho_e
\lesssim 1$, where $\rho_e$ is the thermal electron Larmor radius
\citep{daughton03a}.  For the parameters of our simulations, this can
be written as a condition on the wavelength: $0.5 \lesssim \lambda/d_i
\lesssim 2$.  The fluctuations in the simulation fall within this
range.  On the other hand, the strong, localized electron drift seen
in Figure \ref{jez} differs from systems usually analyzed for the LHDI
instability and the electron Kelvin-Helmholtz instability
\citep{lee15a} is also a possible driver for the turbulence.  Note
that while the instability has reached the non-linear stage by the
time shown in panels (c) and (d), the structure at earlier times (not
shown) exhibits similar spatial scales.  The presence of strong
turbulence around the x-line differs from the results of earlier
three-dimensional simulations, where strong turbulence was largely
localized away from the x-line along the separatrices.

\begin{figure}
\includegraphics[width=0.9\columnwidth]{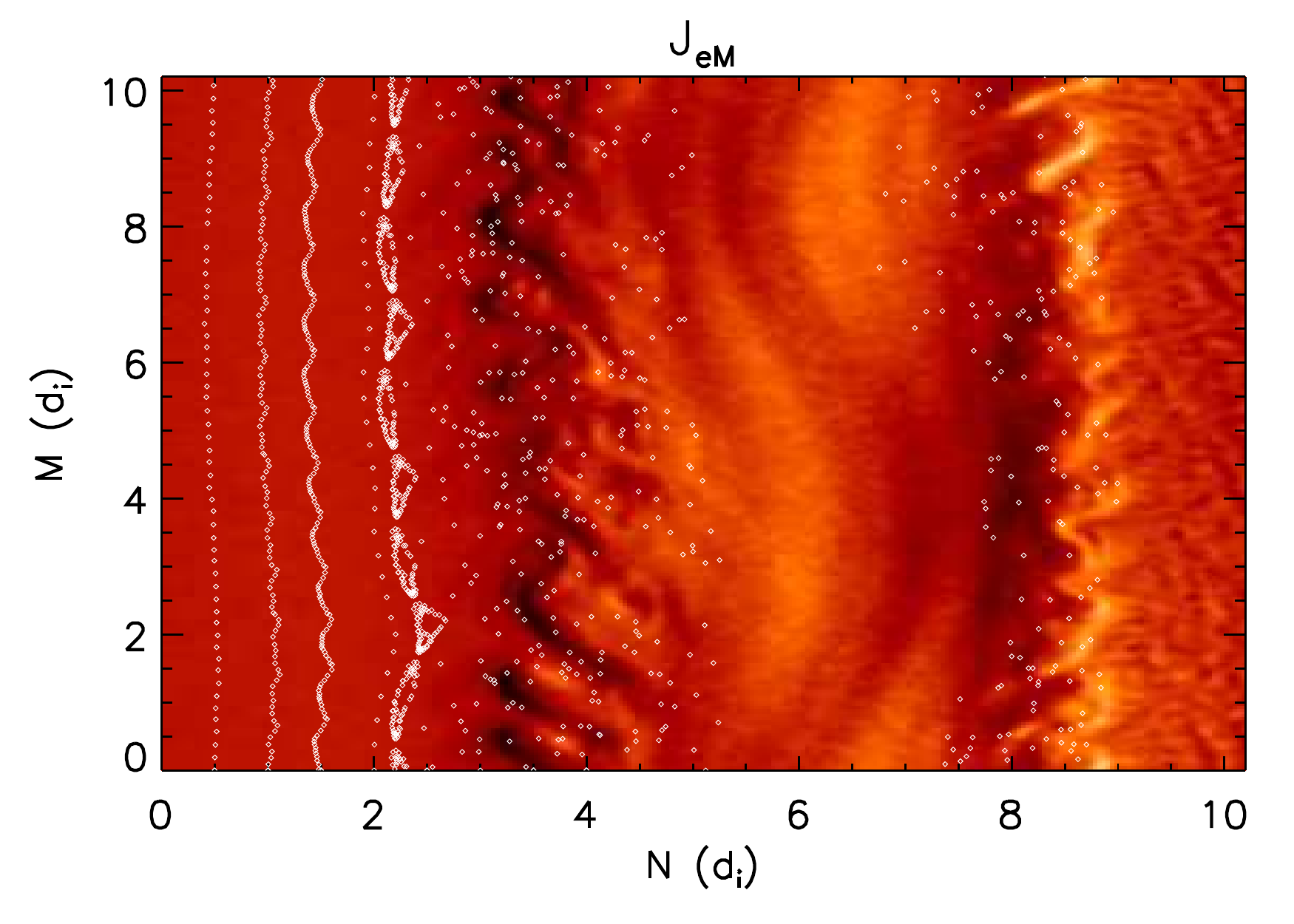}
\caption{\label{phase} Puncture plot showing the intersections of
  field lines with panel (d) of Figure \ref{jez}.  Each dot represents
  the intersection of a field line with the plane after tracing its
  trajectory through the simulation domain.  The islands at $N \approx
  2.5$ mark the transition from laminar to turbulent behavior.}
\end{figure}

The flows driven by the instability are dominantly in the $M-N$ plane
and twist the dominant magnetic field ($L$ direction) so that it
develops $M$ and $N$ components.  We emphasize, however, that the
development of $B_M$ and $B_N$ is a conversion from flow to magnetic
energy rather than the reverse.  Nevertheless, the result is a chaotic
magnetic field. Figure \ref{phase} shows the intersections of field
lines with the $M-N$ plane at the midplane of the magnetic island with
$J_{eM}$ in the background.  Because of the periodic boundary
conditions, each field line passes through the simulation multiple
times, although each pass can also be considered a separate field
line.  On the left side, in the upstream magnetosphere, the field is
laminar.  A band of magnetic flux ropes borders this region, just to
the left of the strongest turbulence which peaks at $N \approx
3.5$. These coherent structures bound the chaotic field lines that
fill the large-scale magnetic island. (The field lines within the
island intersect the plane twice, once at $2.5 \lesssim N \lesssim 5$
and again at $7 \lesssim N \lesssim 8.5$.)  The twisting of flux ropes
by the vortical $M-N$ flows is similar to that inferred from MMS
observations by \cite{ergun16a}.

\begin{figure}
\includegraphics[width=0.9\columnwidth]{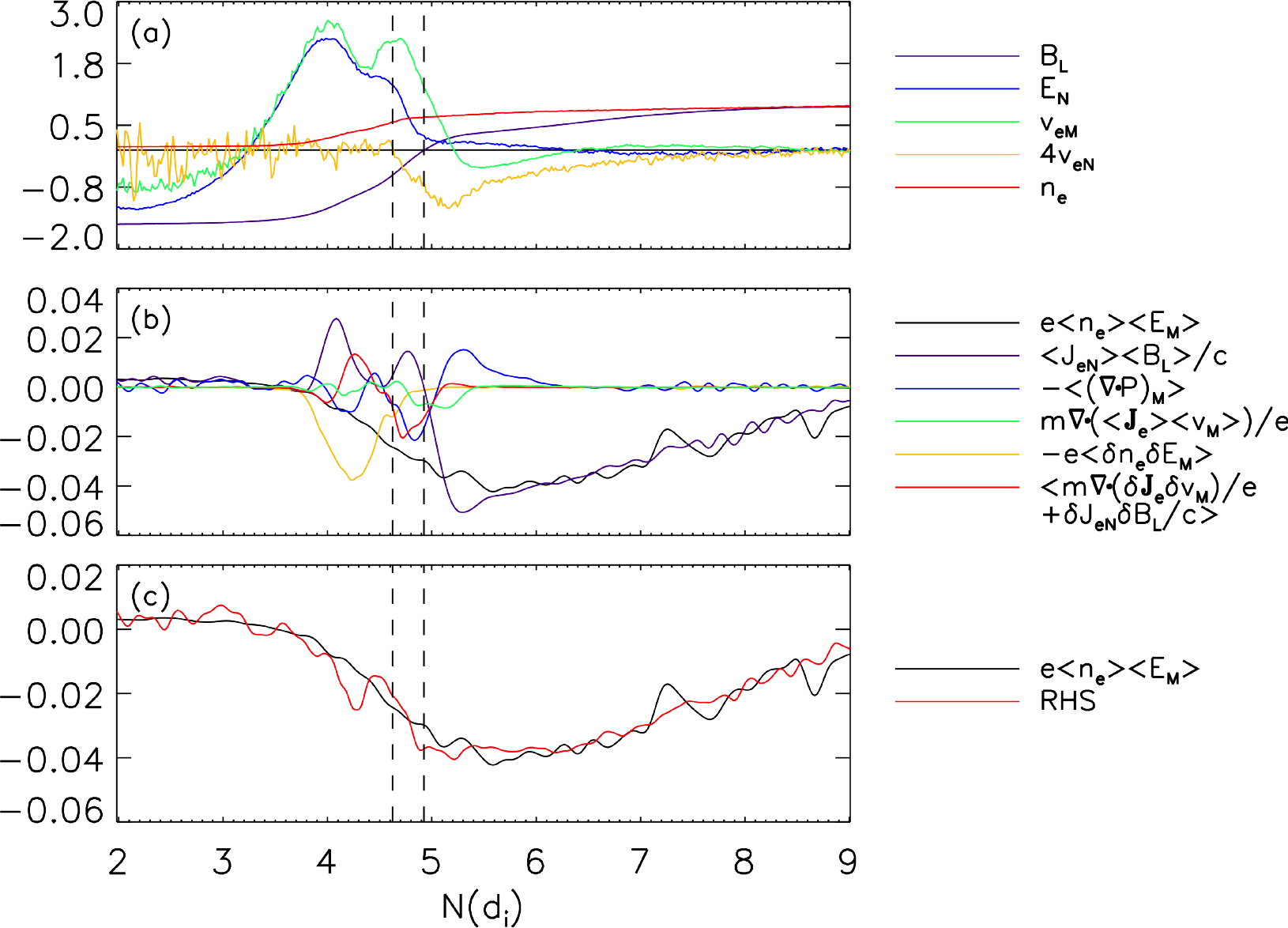}
\caption{\label{ohmplot} Cuts in the $N$ direction through the
  electron diffusion region for the three-dimensional
  simulation. Panel (a): The density $n_e$, reconnecting magnetic
  field $B_L$, the normal electric field $E_N$, and the electron flows
  $v_{eN}$ and $v_{eM}$ all averaged over $M$. Panel (b): The
  principal terms in Ohm's law from equation \ref{ohm3d}.  (Additional
  small terms are included, as noted by the key, but produce minimal
  effects.)  Panel (c): The sum of the left and right sides of
  equation \ref{ohm3d}.  In each panel the vertical lines show the
  approximate positions of the stagnation point ($N\approx 4.5$) and
  x-point ($N\approx 4.9$).}
\end{figure}

The role of turbulence can be quantified by evaluating the terms of
the generalized Ohm's law in a cut through the x-line.  We begin with
the momentum equation for the electron fluid
\begin{equation}\label{ohmeq}
en\mathbf{E} = -mn\frac{d\mathbf{v}}{dt} - \boldsymbol{\nabla
  \cdot}\mathbb{P} - en(\mathbf{v}/c)\boldsymbol{\times}\mathbf{B}
\end{equation}
where $m$, $n$, $\mathbf{v}$, and $\mathbb{P}$ are the electron mass,
density, velocity, and pressure tensor (we only refer to electrons
below and so have dropped the species subscripts).  Taking the
out-of-plane ($M$) component gives, after invoking symmetry with
respect to the $L$ coordinate near the x-line \citep{hesse14a},
\begin{multline}\label{ohm2d}
enE_M = -env_{N}B_L/c - \left(\frac{\partial P_{LM}}{\partial
  L}+\frac{\partial P_{NM}}{\partial N} + \frac{\partial
  P_{MM}}{\partial M}\right) \\ - m\left(\frac{\partial}{\partial t}nv_M
+ \frac{\partial}{\partial N}nv_Nv_M \right)
\end{multline}
In the two-dimensional case $\partial P_{MM}/\partial M = 0$.

In Figure \ref{ohmplot}a we highlight the basics of asymmetric
reconnection by plotting some of the key parameters on a cut along $N$
through the x-line: $n$, $B_L$, $E_N$, $v_{M}$ and $v_{N}$. The
magnetosphere is on the left and the magnetosheath on the right. For
asymmetric reconnection the stagnation point, where $v_{N}=0$, lies on
the magnetosphere side of the x-point, where $B_L=0$
\citep{cassak07a}.  The vertical dashed lines in the figure indicate
the approximate locations of these points. The high-speed electron
flow $v_{M}$ is dominantly driven by $E_N$ and these two quantities
track each other across the diffusion region.  The qualitative
behavior of cuts through the two-dimensional simulation (not shown) is
similar to Figure \ref{ohmplot}a and consistent with the results of
\cite{hesse14a}.  The electron inertia term balances $E_M$ where
$B_L=0$ and the divergence of the pressure tensor balances $E_M$ where
$v_{N}=0$.
 
To establish the role of turbulence in the three-dimensional
simulation, we average over the $M$ direction and decompose every
quantity into a mean and fluctuating component, i.e., $n = \langle n
\rangle + \delta n$. Note that products of quantities produce two
terms, $\langle AB \rangle = \langle A\rangle \langle B\rangle +
\langle \delta A \delta B\rangle$.  Keeping the most significant terms
in equation \ref{ohm2d} gives
\begin{multline}\label{ohm3d}
e\langle n\rangle \langle E_M \rangle = \langle J_N\rangle\langle
B_L\rangle/c - \left[\frac{\partial}{\partial L} \langle P_{LM}\rangle
  + \frac{\partial}{\partial N} \langle P_{NM}\rangle\right] \\+
\frac{m}{e}\left[\frac{\partial}{\partial L} \langle J_L\rangle\langle
  v_M\rangle + \frac{\partial}{\partial N} \langle J_N\rangle\langle
  v_M\rangle\right] \\ -e\langle\delta n\delta E_M\rangle + \left<
\delta J_N\delta B_L/c + \frac{m}{e}\frac{\partial}{\partial N}\delta
J_N\delta v_M \right>
\end{multline}
In deriving equation \ref{ohm3d}, the weak time-dependence has been
dropped since we are focusing on steady-state behavior.  We have also
discarded terms containing $J_L$ and $\delta J_L$ that symmetry
arguments suggest are small (and which we have confirmed are small in
the simulation data).

The first three terms on the right-hand side involve only mean
quantities and can be matched to terms in equation \ref{ohm2d}.  They
represent the usual contributions from the convective motion, pressure
tensor, and inertial terms.  The final two terms arise from the
fluctuations and can be interpreted as contributions from an anomalous
resistivity and an anomalous viscosity associated with the turbulent
transport of the canonical momentum $mv_{M}-eA_M/c$ with $B_L=\partial
A_M/\partial N$, where $\mathbf{A}$ is the vector potential
\citep{che11a}.

Figure \ref{ohmplot}(b) displays the separate terms of equation
\ref{ohm3d} and Figure \ref{ohmplot}(c) shows the left side and the
total of all of the terms on the right side.  (While equation
\ref{ohm3d} includes only the most significant terms, all but the
time-dependent term were kept for the figure.)  The anomalous
resistivity term $\langle\delta n\delta E_M\rangle$ is large around
the stagnation point but diminishes near the x-point while the
viscosity term is significant over a broad region between the two.
Without the inclusion of these terms, the two curves in panel (c)
would not match.  Thus, turbulent effects are playing an essential
role in balancing the reconnection electric field.

Recent investigations of particle distributions in two-dimensional
asymmetric reconnection have revealed crescent-shaped features in the
$v_{M}-v_{N}$ phase space of electrons.  These are signatures of the
cusp-like motion produced by the combination of $E_N$ and a gradient
in $B_L$ \citep{burch16a,bessho16a,shay16a}. If, in the electron
current layers driven by $E_N$, the turbulence is sufficiently strong
the fluctuating electric fields might scatter the electron orbits,
preventing the formation of the crescent distributions.  Of course, if
the electrons were simply gyrating around the $\mathcal{O}(1)$ field,
the turbulence would not strongly affect the orbits unless the
turbulence frequency was comparable to $\Omega_{ce}$. However, instead
the orbits are cusp-like and unmagnetized close to the magnetic null
where they are directly accelerated by $E_N$ across $B_L$
\citep{shay16a,bessho16a}. The motion along $N$ is then turned into
the $M$ direction by $B_L$ to produce the electron drift $v_{eM}$. If
the turbulence breaks up the current layer so that the components of
$E_M$ and $E_N$ are comparable, the electrons will be directly
accelerated in both the $N$ and $M$ directions, potentially disrupting
the cusp-like motions.

\begin{figure}
\includegraphics[width=0.9\columnwidth]{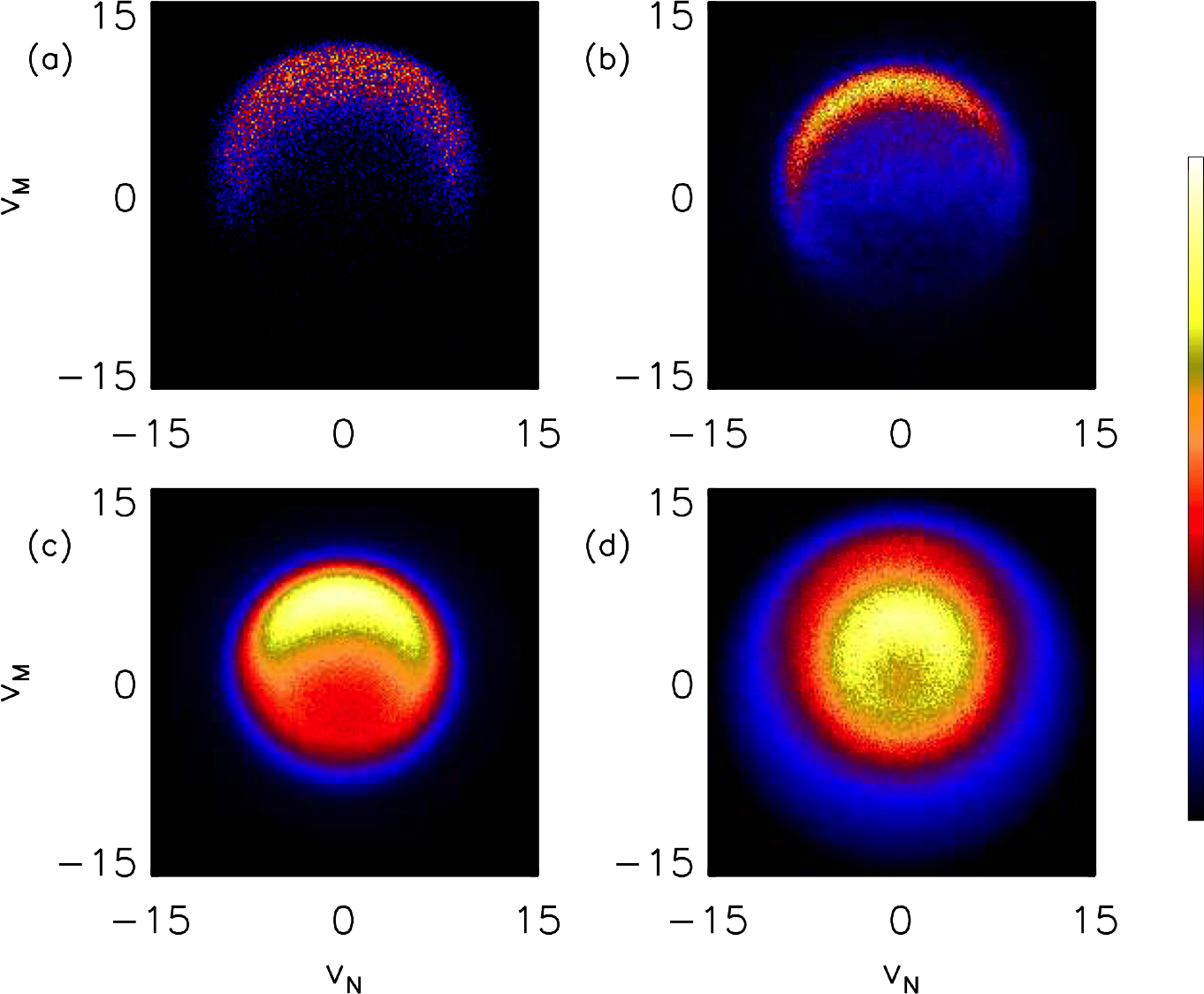}
\caption{\label{cresdf} $v_N-v_M$ electron distribution functions from
  the two-dimensional (panel a) and three-dimensional (panels b-d)
  simulations.  The distributions were taken at the positions shown by
  the stars in Figure \ref{jez}.  Panels b and c were taken near the
  x-line.  In panel b only a limited range in $M$ was sampled, $1 \leq
  M \leq 1.25$; panels c and d sample the entire box, $0\leq M \leq 10.24$.
  Panel d was taken near the separatrix, downstream from the x-line
  (see Figure \ref{jez}).  The number of particles in each velocity
  bin is plotted on a linear scale that is different for each panel,
  although the color bar shows the relative variation.}
\end{figure}

However, Figure \ref{cresdf} suggests that the crescents survive even
when the turbulence in the electron current layers is strong. Panel a
displays data from a region upstream of the x-line on the
magnetosphere side from the two-dimensional simulation.  The crescent
is clearly visible, consistent with earlier simulations
\citep{hesse14a,shay16a,chen16a} and the MMS data \citep{burch16a}.
Data from the three-dimensional simulation, also taken from the
magnetospheric side of the x-line, is shown in panels b and c.  For
panel b the distribution is taken over a limited range in the
out-of-plane direction $1\leq M \leq 1.25$ while panel c is taken over
all $M$.  The crescent is clearly present in panel b.  In panel c,
integration over the larger range in $M$ samples many periods of the
turbulence and smears out, but does not destroy, the crescent.  Panel
d shows a distribution taken near the separatrix but downstream from
the x-line in the three-dimensional simulation.  A crescent feature is
still visible.

The crescents from the two-dimensional and three-dimensional
simulations do exhibit some qualitative differences.  The noisier
distribution of panel a is a consequence of the smaller number of
particles (and hence larger random noise) per velocity bin.  Second,
the two-dimensional case shows a faster bulk flow in the $M$
direction.  This is because the electron current layer in the
two-dimensional case remains highly localized in the $N$ direction.
In contrast, the turbulence in the three-dimensional run broadens the
current layer.  Since the integrated current across the layer must be
the same in both cases, the broader layer from the three-dimensional
run produces a smaller bulk velocity.  On the other hand, the small
counter-clockwise rotation observable in panel b is simply a consequence
of the location at which the distribution is taken.  Similar rotations
can be seen in the two-dimensional simulation for distributions from
nearby locations.

\section{Discussion}\label{discussion}

Reconnection in asymmetric configurations can be stabilized by the
presence of diamagnetic drifts \citep{swisdak03a,swisdak10a,phan13a},
with complete stabilization occurring when the difference in $\beta =
8\pi P/B^2$ between the asymptotic plasmas exceeds $\tan{\theta/2}$,
where $\theta$ is the shear angle between the reconnecting fields.  In
the configuration considered here, $\Delta \beta \approx 2.5$ is
relatively large but, because the guide field is small, $\theta
\approx 170^{\circ}$ is also large.  Hence reconnection is unaffected
by diamagnetic drifts, which is in agreement with the reconnection
rate of $\mathcal{O}(0.1)$ observed for the both the two-dimensional
and three-dimensional simulations.  As a separate effect, a finite
guide field can affect the development of structures in the
out-of-plane direction.  Because $B_M/B_L \lesssim 0.1$ is small in
this case, however, the oblique tearing mode and the development of
flux ropes, as seen in \cite{daughton11a}, does not occur in our
domain.

An important question is whether real mass-ratio simulations would
yield results that differ significantly from the present simulations
where $m_i/m_e=100$. We suggest that the results should not be
sensitive to the mass ratio. Even with real mass ratios the LHDI is
strong in systems with scale lengths near the ion Larmor scale, which
is characteristic of the boundary layers with strong $E_N$ at the
magnetopause. The suppression of LHDI by magnetic shear and finite
$\beta$ is weaker in asymmetric reconnection because the strongest
density gradient and peak current $J_{eM}$, which drive the
instability, are on the magnetosphere side of the x-line where $\beta$
is smaller. The strongest turbulent drag (Figure \ref{ohmplot}(b)) is
peaked near the stagnation point ($v_{eN}=0$), well away from the
magnetic null. The anomalous viscosity terms (Figure \ref{ohmplot}(b))
peak in the region between the magnetic null and the stagnation point
where the gradients in $v_{eM}$ are greatest and have scale lengths
below $d_i$.

In a recent paper \cite{ergun16a} report on MMS observations of very
intense parallel electric fields found in small-scale structures along
the magnetospheric separatrices during magnetopause reconnection. They
associate these parallel electric fields with localized reconnection
events in which the magnetic field is twisted by vortical plasma
motions in the $M-N$ plane. The magnetic turbulence that develops
along the separatrices of our three-dimensional simulations is
reminiscent of these observations -- the strong electron flows
basically twist up the magnetic field. On the other hand, the parallel
electric fields in our simulations are not as intense as in the MMS
data ($\approx 10$ versus $\approx 100$ mV/m) and are largest in the
diffusion region rather than along the separatrices.  Cuts of
$E_{\parallel}$ in the $M-N$ plane through the x-line (not shown)
reveal electron holes similar to those seen in earlier simulations
with larger guide fields \citep{drake03a}.  One possible explanation
for this discrepancy may be the artificially low mass ratio.  A
realistic value could yield sharper gradients and more intense fields.
It is also possible that in our simulations we are only exploring the
early stages of the dynamics of these turbulent current layers. With
larger simulations that could be evolved for longer times it is
possible that the strong parallel currents that develop along the
separatrices might form more intense localized parallel electric
fields as seen in some earlier two-dimensional simulations
\citep{cattell05a,lapenta11a}.

The role that turbulence might have in breaking the frozen-in
condition has not yet been explored with the MMS data. On the other
hand, short bursts of $E_M\sim 10$ mV/m were seen in the current layer
where $E_N$ is large \citep{burch16a}. Thus, the presence of
turbulence seems likely but its consequences and the specific
correlated averages that need to be carried out to evaluate the
anomalous drag and viscosity coefficients in equation \ref{ohm3d} have
not been evaluated.

In conclusion, we find that the inclusion of the third dimension
permits the development of strong turbulence, both at the x-line and
along the separatrices.  This turbulence makes significant
contributions to the balance of Ohm's law but, perhaps surprisingly,
does not disrupt the formation of crescent features in the velocity
distribution functions.  Hence, the existence of such crescents cannot
serve as an indicator as to whether turbulence plays an important role
at a reconnection x-line.

\begin{acknowledgments}
This work was supported by NASA grants NNX14AC78G, NNX16AG76G, and
NNX16AF75G and NSF grants PHY1500460, AGS-0953463, and
AGS-1460037. The simulations were carried out at the National Energy
Research Scientific Computing Center.  The data used to perform the
analysis and construct the figures for this paper are available upon
request.

\end{acknowledgments}

%\end{article}

%\bibliographystyle{agu04}
%\bibliography{paper}

\end{document}